\documentclass[twocolumn,superscriptaddress,showpacs,preprintnumbers,amsmath,amssymb,oneside,prl]{revtex4}
\usepackage{graphicx}
\usepackage{dcolumn}
\usepackage{bm}

\begin{document}
\title{Experimental Free-Space Distribution of Entangled Photon
Pairs over \\ a Noisy Ground Atmosphere of 13km}% Force line breaks with \\
\author{Cheng-Zhi Peng}
\affiliation{Department of Modern Physics and Hefei National
Laboratory for Physical Sciences at Microscale, University of
Science and Technology of China, Hefei, Anhui 230026, China}
\author{Tao Yang}
\affiliation{Department of Modern Physics and Hefei National
Laboratory for Physical Sciences at Microscale, University of
Science and Technology of China, Hefei, Anhui 230026, China}
\author{Xiao-Hui Bao}
\affiliation{Department of Modern Physics and Hefei National
Laboratory for Physical Sciences at Microscale, University of
Science and Technology of China, Hefei, Anhui 230026, China}
\author{Jun-Zhang}
\affiliation{Department of Modern Physics and Hefei National
Laboratory for Physical Sciences at Microscale, University of
Science and Technology of China, Hefei, Anhui 230026, China}
\author{Xian-Min Jin}
\affiliation{Department of Modern Physics and Hefei National
Laboratory for Physical Sciences at Microscale, University of
Science and Technology of China, Hefei, Anhui 230026, China}
\author{Fa-Yong Feng}
\affiliation{Department of Modern Physics and Hefei National
Laboratory for Physical Sciences at Microscale, University of
Science and Technology of China, Hefei, Anhui 230026, China}
\author{Bin Yang}
\affiliation{Department of Modern Physics and Hefei National
Laboratory for Physical Sciences at Microscale, University of
Science and Technology of China, Hefei, Anhui 230026, China}
\author{Jian Yang}
\affiliation{Department of Modern Physics and Hefei National
Laboratory for Physical Sciences at Microscale, University of
Science and Technology of China, Hefei, Anhui 230026, China}
\author{Juan Yin}
\affiliation{Department of Modern Physics and Hefei National
Laboratory for Physical Sciences at Microscale, University of
Science and Technology of China, Hefei, Anhui 230026, China}
\author{Qiang Zhang}
\affiliation{Department of Modern Physics and Hefei National
Laboratory for Physical Sciences at Microscale, University of
Science and Technology of China, Hefei, Anhui 230026, China}
\author{Nan Li}
\affiliation{Department of Modern Physics and Hefei National
Laboratory for Physical Sciences at Microscale, University of
Science and Technology of China, Hefei, Anhui 230026, China}
\author{Bao-Li Tian}
\affiliation{Department of Modern Physics and Hefei National
Laboratory for Physical Sciences at Microscale, University of
Science and Technology of China, Hefei, Anhui 230026, China}
\author{Jian-Wei Pan}
\affiliation{Department of Modern Physics and Hefei National
Laboratory for Physical Sciences at Microscale, University of
Science and Technology of China, Hefei, Anhui 230026, China}
\affiliation{Physikalisches Institut der Universitaet Heidelberg,
Philosophenweg 12, Heidelberg 69120, Germany}

\date{\today}% It is always \today, today,

\begin{abstract}
We report free-space distribution of entangled photon pairs over a
noisy ground atmosphere of 13km. It is shown that the desired
entanglement can still survive after the two entangled photons
have passed through the noisy ground atmosphere. This is confirmed
by observing a space-like separated violation of Bell inequality
of $2.45 \pm 0.09$. On this basis, we exploit the distributed
entangled photon source to demonstrate the BB84 quantum
cryptography scheme. The distribution distance of entangled photon
pairs achieved in the experiment is for the first time well beyond
the effective thickness of the aerosphere, hence presenting a
significant step towards satellite-based global quantum
communication.
\end{abstract}


\maketitle

In the future large scale realization of quantum communication
schemes \cite{bb84,ekert91,bennett93}, we have to solve the
problems caused by the photon loss and decoherence in the
transmission channel. For example, because of the photon loss and
the unavoidable dark count of the current available single-photon
detectors£¬the maximum distance in the fibre-based quantum
cryptography is limited to the order of 100km \cite{gisin_rmp}.
The quantum repeater scheme that combines entanglement swapping,
entanglement purification and¡¡quantum memory
\cite{repeater,swapping,purification} proposed an efficient way to
generate highly entangled states between distant locations, hence
providing an elegant solution to the photon loss and decoherence
problem. In recent years, significant progress has been achieved
in the experimental demonstration of entanglement swapping,
entanglement purification and quantum memory
\cite{swapping_exp,purification_exp,atom_nature,memory_sci,matter_light},
yet one still has long way to go before the above techniques can
be finally integrated into a single unit in order to be useful for
realistic quantum communication over large distances.

Another promising way out to realize long-distance quantum
communication is to exploit satellite-based free-space
distribution of single photons or entangled photon pairs
\cite{satellite}. In the scheme, the photonic quantum states are
first sent through the aerosphere, then reflected from one
satellite to another and finally sent back to the earth. Since the
effective thickness of the aerosphere is on the order of 5-10km
(i.e. the whole aerosphere is equivalent to 5-10km ground
atmosphere) and in the outer space the photon loss and decoherence
is negligible, with the help of satellites one can achieve global
free-space quantum communication as long as the quantum states can
still survive after penetrating the aerosphere \cite{satellite}.

Along this line, important experimental progress has been made
very recently in the free-space distribution of attenuated laser
pulses and entangled photon pairs \cite{free_23km,free_600m}.
However, on the one hand, in the quantum cryptography experiment
with attenuated laser pulses \cite{free_23km} the huge photon loss
in the transmission channel leaves an eavesdropping loophole. This
is because the eavesdropper could in principle exploit a
transmission channel with less or without photon loss and only
allow those attenuated laser pulses containing two or more photons
to reach the receiver. In this way, the eavesdropper can use a
beamsplitter to steal at least one photon from these specific
attenuated laser pulses without being detected. On the other hand,
while the achieved distance in the previous entanglement
distribution experiment \cite{free_600m} is only on the order of
600m which is far below the effective thickness of the aerosphere,
the achieved low transmission efficiency ($\sim10^{-3}$) would not
enable a sufficient link efficiency over large distances, which
is, however, required for satellite-based free-space quantum
communication \cite{satellite}.

In this letter, with the help of laser-pulse-assisted
synchronization method and own designed telescope systems we drive
the free-space technology further by reporting free-space
distribution of entangled photon pairs over a noisy ground
atmosphere of 13km. We confirm that the desired entanglement can
still survive after the two entangled photons have passed through
the noisy ground atmosphere with a distance beyond the effective
thickness of the aerosphere. In addition, we also exploit the
distributed entangled photon source to experimentally demonstrate
the BB84 quantum cryptography scheme \cite{bb84} without the
eavesdropping loophole. The distribution distance of entangled
photon pairs achieved in our experiment is for the first time well
beyond the effective thickness of the aerosphere, hence presenting
a significant step towards satellite-based global quantum
communication.

\begin{figure}[ptb]
\includegraphics[width=\columnwidth]{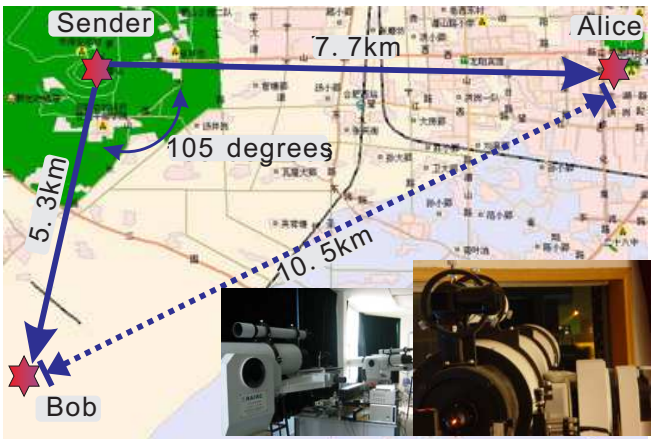}
\caption{Schematic diagram of locations in our experiment. The
source of entangled photons is located at the foot of a high
television tower, on the top of Dashu Mountain. Alice is located
on the west campus of USTC, and Bob is located at Feixi, a county
of Hefei city. Photons from the Sender to the receivers experience
noisy city environment. Therefore, strongly influenced by the air
pollution and noisy background lights, even after rain with less
air pollution, the background count rate can reach about 30,000
per second at night without using interference filters.}
\label{fig1}
\end{figure}

In the experiment, as shown in Fig.\ref{fig1}, a Sender is located
on the top of Dashu Mountain in Hefei of China, with an elevation
of 281m, and two receivers (Alice and Bob) are located at the west
campus of USTC and at Feixi of Hefei respectively. The direct
distance between the two receivers is about 10.5km. The distances
from the Sender to Alice and from the Sender to Bob are, 7.7km and
5.3km, respectively. One of the two entangled photons passes
through nearly half part of Hefei city, experiencing an extremely
challenging environment above the city. The two receivers are not
in sight with each other due to the existence of many buildings
between them.

At the Sender, we utilize type-II parametric down-conversion to
generate entangled photon pairs \cite{kwiat95}. The Argon Ion
laser used to pump BBO crystal has a wavelength of 351nm. When the
pump power is about 300mw, with narrow bandwidth filters of 2.8nm
in front of single-photon detectors we locally collect about
10,000 pairs of entangled photons per second, with an average
single photon count rate about 155,000 per second.

In order to optimize the transmission efficiency and its
stability, we have designed two sets of transmission system
ourselves, which contain 4 large telescopes of refraction type.
Each telescope weighs about 800kg, with a focus length about 2m.
Lenses used in the telescope are coated to maximize the peak
transmission rate at the wavelength of entangled photons (702nm).
Each transmission system composed of two telescopes can achieve a
transmission rate of above $70\%$. The wild environment on the top
of Dashu Mountain brings us many difficulties, and we have taken
several measures to overcome them. The two telescopes at the
sender are specially designed to be robust against the strong
wind.

The entangled photons at the Sender are collected into two
single-mode fibers, which are connected to the two sending
telescopes respectively. Due to the disturbance of the atmosphere,
the size and position of the received beam vary randomly, causing
reduction of the collecting efficiency. To solve this problem, we
have used the two sending telescopes to expand the beam diameter
to about 12cm for long-distance propagation. Moreover, at each
receiver a similar telescope is used to receive the entangled
photons. After being focused, entangled photons are coupled into
62.5um multi-mode fibers and finally sent to single photon
detectors. With these efforts, we manage to keep the transmission
system to work stably for a couple of hours. For example, in the
right photography of Fig.\ref{fig1} we can see at the Alice side a
bright and stable adjusting laser beam from the Sender.

\begin{figure}[ptb]
\includegraphics[width=\columnwidth]{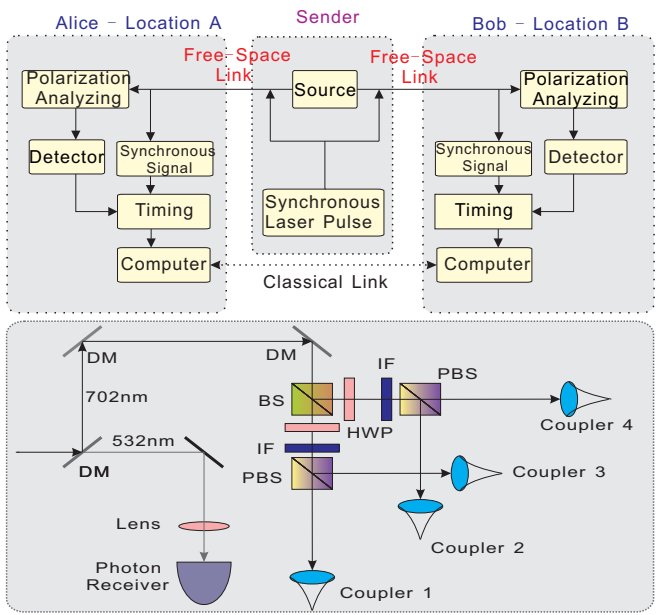}
\caption{Block diagram of the experiment and optical setup at the
receivers. As shown in the left Figure, we combine the entangled
photons with the synchronous pulsed laser beam utilizing a
dichroic mirror (DM) at the Sender and separate them at each
receiver with a DM. Then it follows with single photon
polarization analysis and time synchronization. In the down
figure, a beam splitter (BS) is used to achieve random basis
selection, and a half-wave plate (HWP) together with a
polarization beam splitter (PBS) make up an apparatus for
polarization measurement. Interference filters (IF) are used to
get rid of noisy background light.} \label{fig2}
\end{figure}

Since the distances from Sender to Alice and to Bob are not equal,
the two entangled photons will arrive at each receiver at a
different time. The air disturbance will cause this time
difference varying randomly, result in a time difference shake
($\Delta T$). To coincide the detected events at the two
receivers, we have to make sure that the coincident time window
should be wider than ($\Delta T$). However, when we widen the
coincident time window to get the adequate true coincident events,
the accidental coincident count rate also increases and thus
results in a reduction of the visibility. In our experiment, we
utilized the method of laser pulse synchronization to achieve time
coincidence between the two receivers (See Fig.\ref{fig2}). At the
Sender, Q-switched laser pulses with a wavelength of 532nm are
separated into two parts, and then sent to the receivers,
experiencing the same optical path as the entangled photons. At
each receiver, we measure the time difference between the signal
of single-photon event and the signal of the corresponding
synchronous laser pulse for subsequent coincidence via classical
communication link. Consider other ingredients causing time shake,
we set the time window to 20ns in our experiment.

Finally, to minimize background count rate, 2.8nm interference
filters are utilized at each receiver to block the noisy
background light. With the filters added, the average background
count rate is about 400 per second. When the weather condition is
perfect with a considerable high vis. ($>15$km), total single
photon count rate is about 40,000 per second at Bob, about 18,000
per second at Alice, and the coincident count rate is about 300
per second. At the normal vis. (~10km), the coincident count rate
is about 150 per second.

\begin{figure}[ptb]
\includegraphics[width=\columnwidth]{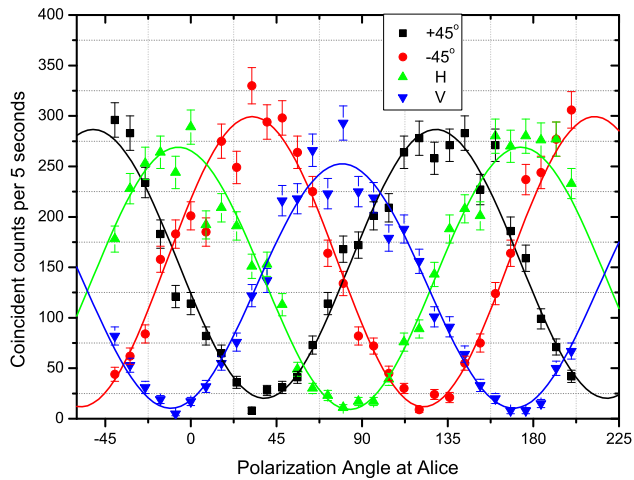}
\caption{Verification of the distributed quantum entanglement. In
order to test the quality of the entangled state between the two
receivers, we measured the coincident count as a function of
Alice's polarization angle. Four curves correspond to four
polarization angles (H,V,+45, -45) set at Bob. The data is best
fitted with sin functions, showing that the visibility is 94\% in
H/V basis and it is 89\% in +45/-45 basis. The average visibility
has reached 91\%, which is far beyond the threshold required for a
violation of Bell inequality.} \label{fig3}
\end{figure}

The entangled state prepared at the Sender can be expressed as
follows,
\begin{equation}
|\psi^-\rangle=\frac{1}{\sqrt2}(|H\rangle_A|V\rangle_B-|V\rangle_A|H\rangle_B)
\end{equation}
where photon A is sent to Alice, photon B is sent to Bob, H and V
represent horizontal and vertical polarization. The local
visibility at the Sender is about 98\% in the H/V basis, and 94\%
in the +45/-45 basis. The observed visibilities between two
separated receivers are 94\% and 89\% in the H/V and +45/-45
bases, respectively (see Fig.\ref{fig3}). Hence the average
visibility reaches 91\%, which is well beyond 71\% required for a
violation of Bell inequality. In order to further test the quality
of the entangled state, we measured the Clauser-Horne-Shimony-Holt
(CHSH) inequality which is one type of Bell inequalities
\cite{chsh}. The polarization correlation coefficient is defined
as follows,
\begin{equation}
E(\phi_A,\phi_B)=\frac{N_{++}+N_{--}-N_{+-}-N_{-+}}{N_{++}+N_{--}+N_{+-}+N_{-+}}
\end{equation}
Where $N_{ij}(\phi_A,\phi_B)$ are the coincidences between the $i$
channel of the polarizer of Alice set at angle $\phi_A$  and the
$j$ channel of the polarizer of Bob set at angle $\phi_B$. In the
CHSH inequality, parameter $S$ is defined as,
\begin{equation}
S=|E(\phi_A,\phi_B)-E(\phi_A,\phi_B')+E(\phi_A',\phi_B)+E(\phi_A',\phi_B')|
\end{equation}
In the local realistic view, no matter what angles $\phi_A$ and
$\phi_A$ are set to, parameter S should be below 2. But in the
view of quantum mechanics, $S$ will get to the maximal value
$2\sqrt2$ when the polarization angles are set to
$(\phi_A,\phi_A',\phi_B,\phi_B')=(0^\circ,45^\circ,22.5^\circ,67.5^\circ)$

\begin{table}
\centering \caption{Measured correlation coefficients required for
CHSH inequality} \label{table1}
\begin{tabular*}{1\columnwidth}{@{\extracolsep{\fill}}ccccc}
\hline \hline
$E(\phi_A,\phi_B)$ & $(0^\circ,22.5^\circ)$ & $(0^\circ,67.5^\circ)$ & $(45^\circ,22.5^\circ)$ & $(45^\circ,67.5^\circ)$\\
\hline \hline
Value & -0.681 & 0.764 & -0.421 & -0.581\\
\hline
Deviation & 0.040 & 0.036 & 0.052 & 0.046 \\
\hline
\end{tabular*}
\end{table}

As the detection loophole existed in all previous photonic test of
local realism
\cite{bell_test_first,bell_test_zeilinger,bell_test_gisin}, here
we are only going to show a violation of Bell inequality with
space-like separated observers. To do so, at each observer a
beamsplitter (see Fig.\ref{fig2}) is used to achieve the true
random basis selection and use 4 single photon detectors to
perform the $(0^\circ,45^\circ,22.5^\circ,67.5^\circ)$
polarization measurement. With emphasis, we note that our passive
beamsplitter is sufficient to provide a bona fide test of the
locality loophole. This is because in experiments using low
efficient detectors the locality loophole can be closed
equivalently using active or passive switches
\cite{active_passive}. In our experiment, the whole measurement
progress was completed in 20 seconds. Note that, all the 16
coincident counts are measured simultaneously with two receivers
space-like separated. Moreover, since different detectors have
different detection efficiency two-fold coincidence normalization
has been performed based on the single count rate. The measured
result of parameter S is $2.45\pm0.09$, with a violation of the
CHSH inequality by 5 standard deviations (see Table \ref{table1}
for details). This result firmly ascertains that entanglement has
been built between the two distant receivers.

In the cryptography experiment, we take a variant scheme of BB84
with entangled photons \cite{bb84}. In this scheme, with the help
of the beamsplitter Alice and Bob randomly measure her/his
received photons in the H/V or +45/-45 basis. Because of the
perpendicular property in polarization of $|\psi^-\rangle$ , when
they have chosen the same basis, their private keys are
anti-correlated. Then identical keys can be easily got, if one of
them converts all her/his keys. In 4 minutes, we obtained 29,433
coincidence events. Due to the difference of the collecting
efficiency of the four couplers at each receiver, we have randomly
discarded some events related to the high-efficiency couplers in
order to make the efficiency of the couplers at each receiver
equal, and after this progress, we got 15,308 coincidence events.
Discarding the events that Alice and Bob had chosen different
bases, we got 7,956 bits of sifted key, and the QBER is 5.83\%.
Then we did error correction, and decreased the key size to 4,869
bits with an error rate of 1.46\%. After the privacy amplification
procedure depending on the QBER, finally we got 2435 bits of final
secure key. This corresponds an average key distribution rate of
10bit/second. Note that, using a recent high-intensity entangled
photon source \cite{high_intensity} one can easily increase the
average key distribution rate to a few hundreds per second.

Although compared to the previous experiments our experimental
results might seem to be only a modest step forward, the
implication is profound. First, our experiment demonstrated for
the first time the entanglement can still survive after
penetrating the effective thickness of the aerosphere by showing a
violation of Bell inequality with space-like separated observers.
Obviously, the strong violation of Bell inequality is sufficient
to guarantee the absolute security of the quantum cryptography
scheme, hence closing the eavesdropping loophole. Second, the link
efficiency of entangled photon pairs achieved in our experiment is
about a few percent, which is well beyond the threshold required
for satellite-based free-space quantum communication
\cite{satellite}. Finally, the methods developed in the present
experiment to establish a high stable transmission channel and
achieve synchronization between two distant receivers provide the
necessary technology for future experimental investigations of
global quantum cryptography and quantum teleportation in
free-space.

This work was supported by the National Natural Science Foundation
of China, Chinese Academy of Sciences and the National Fundamental
Research Program. This work was also supported by the Alexander
von Humboldt Foundation and the Marie Curie Excellence grant of
the European Commission. The authors would like to thank the great
help of Anhui Television Station.

\end{document}